\documentclass[%preprint,
superscriptaddress,
 amsmath,amssymb,
 aps,
reprint
]{revtex4-2}

\usepackage{color}
\usepackage{graphicx}% Include figure files
\usepackage{dcolumn}% Align table columns on decimal point
\usepackage{bm}% bold math
\usepackage{epstopdf}
\usepackage{gensymb}
\usepackage{hyperref}
\usepackage{natbib}
\usepackage{float}

\begin{document}

%\preprint{APS/123-QED}

\title{Strain patterning of flexomagnetism}% Force line breaks with \\
%\thanks{A footnote to the article title}%

\author{Tamalika Samanta}
\affiliation{Materials Science and Engineering, University of Wisconsin-Madison, Madison, WI 53706, USA}

\author{Zachary T. LaDuca}
\affiliation{Materials Science and Engineering, University of Wisconsin-Madison, Madison, WI 53706, USA}

\author{An-Hsi Chen}
\affiliation{Materials Science and Technology Division, Oak Ridge National Laboratory, Oak Ridge, TN, 37831, USA}

\author{Sangsoo Kim}
\affiliation{Materials Science and Technology Division, Oak Ridge National Laboratory, Oak Ridge, TN, 37831, USA}

\author{Ying-Ting Chan}
\affiliation{Department of Physics and Astronomy, Rutgers University, Piscataway, NJ 08854, USA}

\author{Jiaxuan Wu}
\affiliation{Materials Science and Engineering, University of Wisconsin-Madison, Madison, WI 53706, USA}

\author{Yujia Teng}
\affiliation{Department of Physics and Astronomy, Rutgers University, Piscataway, NJ 08854, USA}

\author{Debarghya Mallick}
\affiliation{Materials Science and Technology Division, Oak Ridge National Laboratory, Oak Ridge, TN, 37831, USA}

\author{Matthew Brahlek}
\affiliation{Materials Science and Technology Division, Oak Ridge National Laboratory, Oak Ridge, TN, 37831, USA}

\author{T. Zac Ward}
\affiliation{Materials Science and Technology Division, Oak Ridge National Laboratory, Oak Ridge, TN, 37831, USA}

\author{Katherine Su}
\affiliation{Materials Science and Engineering, University of Wisconsin-Madison, Madison, WI 53706, USA}

% \author{Rui Liu}
% \affiliation{Advanced Photon Source, Argonne National Lab, Lemont, IL}

\author{Jia-Mian Hu}
\affiliation{Materials Science and Engineering, University of Wisconsin-Madison, Madison, WI 53706, USA}

\author{Weida Wu}
\affiliation{Department of Physics and Astronomy, Rutgers University, Piscataway, NJ 08854, USA}

\author{Turan Birol}
\affiliation{Department of Chemical Engineering and Materials Science, University of Minnesota, Twin Cities, Minneapolis, MN, USA}

\author{Hanfei Yan}
\affiliation{National Synchrotron Light Source II, Brookhaven National Laboratory, Upton, New York, USA}

\author{Michael S. Arnold}
\affiliation{Materials Science and Engineering, University of Wisconsin-Madison, Madison, WI 53706, USA}

\author{Karin M. Rabe}
\affiliation{Department of Physics and Astronomy, Rutgers University, Piscataway, NJ 08854, USA}

\author{Jason K. Kawasaki} \email{Author to whom correspondence should be addressed: Jason Kawasaki, jkawasaki@wisc.edu}
\affiliation{Materials Science and Engineering, University of Wisconsin-Madison, Madison, WI 53706, USA}

\date{\today}% It is always \today, today,
             %  but any date may be explicitly specified
\begin{abstract}
Flexomagnetism, the coupling of magnetic ordering to strain gradients, provides access to novel symmetry-broken magnetic phases that cannot be accessed via uniform strain.
%Flexomagnetism is the coupling of magnetism to elastic strain gradients, which break inversion symmetry and enable Dzyaloshinskii–Moriya interactions and chiral spin textures in otherwise centrosymmetric materials. These effects are unattainable under uniform strain alone. 
However, flexomagnetism is hard to understand because it is extremely difficult to control a spatially varying strain. Here, we develop a top-down strategy to pattern transverse strain gradients using helium ion implantation through a lithographically defined mask. Using epitaxial films of the antiferromagnetic nodal line semimetal GdAuGe, we demonstrate that transverse strain gradients $\partial \varepsilon_{zz}/\partial x$ induce near-room-temperature ferromagnetic response, compared to the retained para or antiferromagnetism for homogeneously strained GdAuGe. We spatially correlate the magnetic response with the regions of largest strain gradient, via magnetic force microscopy and nanobeam x-ray diffraction, respectively, to confirm the flexomagnetic response. %\textcolor{red}{constrain the form of the flexomagnetic tensor [update this based on level of theory?]}. 
Our approach opens new avenues for the precise control of magnetic phases in thin films of quantum materials via a patterned strain gradient.

%is significant because, unlike the magnetostrictive coupling to homogeneous strain, flexomagnetism is an odd function of the strain gradient that enables bistable switching of the sign of magnetization. By combining magnetometry and nanobeam x-ray diffraction we were able to map the strain profile and have a better understanding of the couplings between strain gradient, magnetic anisotropy, and exchange interactions in this system. 

\end{abstract}

\maketitle

%\section*{Introduction}
Strain gradients are higher order lattice distortions that differ fundamentally from uniform strain because they break inversion symmetry and activate magnetic order parameters that are absent in centrosymmetric structures\cite{makushko2022flexomagnetism, eliseev2009spontaneous, liu2024flexomagnetoelectric}, e.g. Dzyaloshinskii-Moriya Interaction (DMI)-induced chiral spin textures. Through their coupling with magnetization ($M$), known broadly as flexomagnetism, strain gradients can alter exchange interactions, magnetic anisotropy, domain-wall energetics, and magnetic transition temperatures\cite{ahadi2019enhancing}. This sensitivity makes them a powerful, yet underexplored tuning parameter for controlling magnetism in quantum materials.

From a theoretical standpoint, flexomagnetism remains challenging to treat within a unified microscopic framework that consistently captures exchange interactions, magnetic anisotropy, and Dzyaloshinskii–Moriya interactions under spatially varying strain. Explicitly incorporating strain gradients into first-principles density functional theory calculations generally requires prohibitively large supercells, rendering such approaches difficult \cite{qiu2023flexo,qiao2024curvature,tang2025intrinsic}. As a result, current theoretical understanding relies predominantly on phenomenological models, which are typically parameterized for specific materials and geometries\cite{gong2025large}.
%These models are highly material-specific and not easily generalized. 
%\textcolor{red}{These models require empirical fitting and thus it is difficult to make materials-specific prediction (?)} 
Well controlled strain tensor and experiments are needed to bridge the gaps between atomic scale exchange and mesoscale phenomenology. 

\begin{figure}[t]
    \centering
    \includegraphics[width=0.4\textwidth]{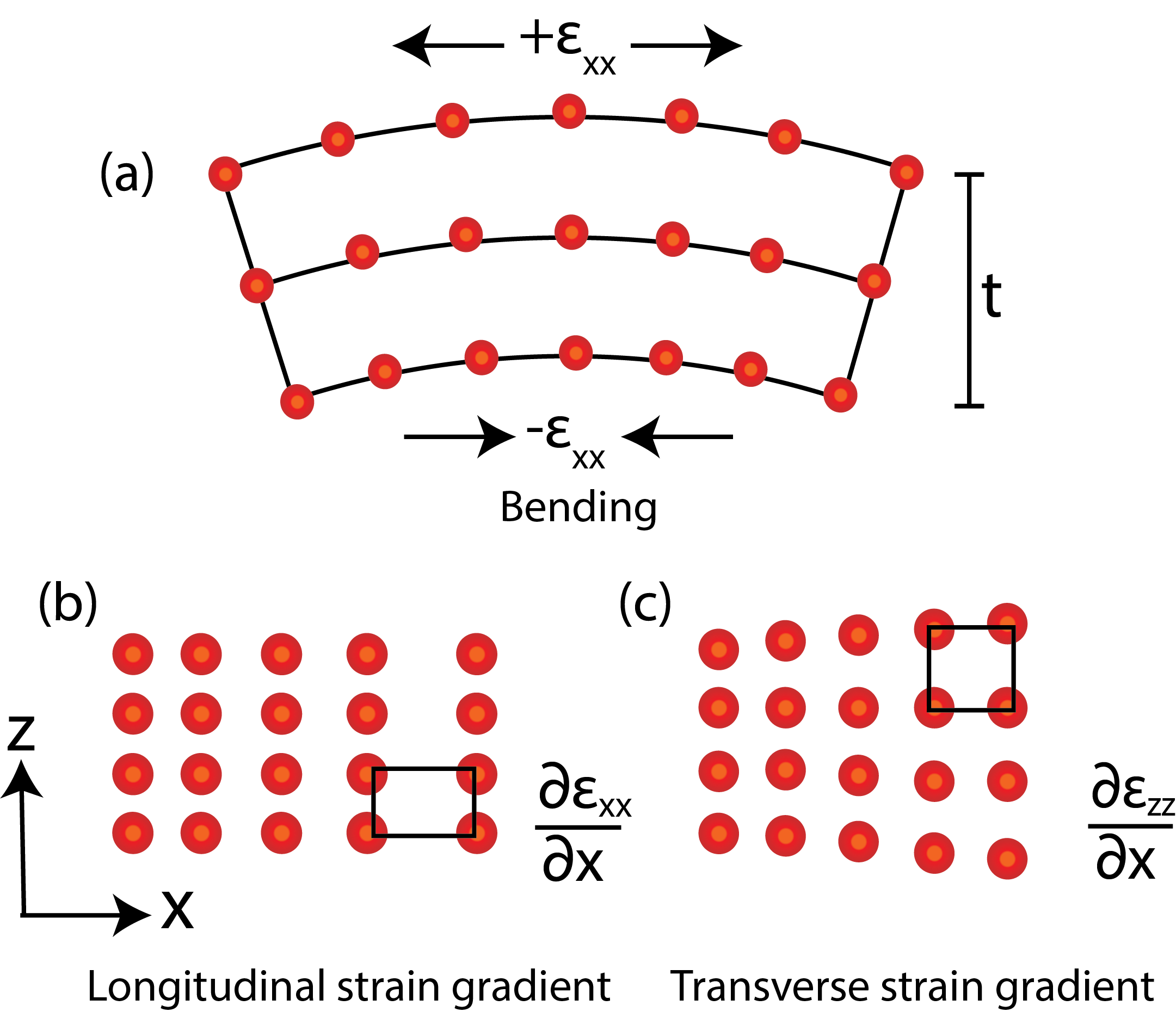}
    \caption{\textbf{Types of strain gradients.} (a) Bending induced strain gradients. The $\pm$ $\varepsilon_{xx}$ refers to the tensile and compressive strain present in the lattice in a bent geometry which can be considered as half of a single wrinkle. In such geometry both transverse strain gradient and longitudinal strain gradient occur. Break down of the dominant strain gradient components: (b) longitudinal, $\partial \varepsilon_{xx} /\partial x$ and (c) transverse, $\partial \varepsilon_{zz} /\partial x$. The intrinsic longitudinal strain gradient resembles an inhomogeneous uniaxial deformation, while the transverse component reflects in-plane spatial variations of the transverse normal strain. Shear component of strain is not shown.}
\label{strain_comp}
\end{figure}

Yet experimentally it remains a challenge to control components of strain tensor\cite{kabychenkov2019flexomagnetic}. 
%as the strain gradient field with multiple independent components, each couples to magnetization in a symmetry-dependent manner\cite{kabychenkov2019flexomagnetic}.
%The most common method to achieve large flexomagnetic responses on the order of $10^{6}-10^{8}$\,m$^{-1}$ is applying mechanical instabilities such as bending in the form of wrinkling or corrugation \cite{du2021epitaxy,ling2023flexomagnetic,chizhik2025direct,harbola2021strain}. 
The most common method is bending, which can produce strain gradients as large as $10^{6}-10^{8}$\,m$^{-1}$ \cite{du2021epitaxy,ling2023flexomagnetic,chizhik2025direct,harbola2021strain} in natively van der Waals materials \cite{kang2021pseudo, kim2024stretchable} and in released epitaxial membranes of oxides and intermetallics \cite{Daesu2020flexoelectricity,du2023strain, du2021epitaxy, laduca2024cold,harbola2021strain}.
%Such bending geometries have been exploited in oxides and intermetallic films to tune electronic and magnetic responses\cite{Daesu2020flexoelectricity,du2024tunable}. 
However, these approaches face fundamental limitations. First, only ultrathin membranes of three-dimensional materials or flexible two-dimensional materials can sustain the large elastic bending deformations needed to achieve significant gradients\cite{liu2022flexoresponses,edstrom2022curved,ga2022dzyaloshinskii,shuaizhao2025straingradient}. Second, and more critically, large bending deformations introduce a complicated strain gradient profile. This can be easily understood from Fig.\ref{strain_comp}. The typical bending type deformation introduces both longitudinal ($\partial\varepsilon_{xx}/\partial x$) and transverse ($\partial\varepsilon_{zz}/\partial x$) strain gradients, as well as localized shear near inflection points, so the lattice rarely experiences a single, well-defined strain gradient component. It is therefore difficult to isolate a specific flexomagnetic channel or extract intrinsic coupling coefficients as the strain gradient couples to magnetization in a symmetry-dependent manner. Compared to bending, partially relaxed epitaxial films can, in principle, generate more directional gradients (~$10^{6}$\,m$^{-1}$), in the form of $\partial\varepsilon_{zz}/\partial z$\cite{zhang2021strain} or $\partial\varepsilon_{xx}/\partial z$\cite{makushko2022flexomagnetism}. However, this approach is not ideal as such strain gradients is achieved through strain relaxation and can be irregular, hard to control, and unstable because of dislocation motion and clamping of the substrate.

Here, we demonstrate a method for top-down patterning of transverse strain gradients. %We utilize Helium implantation which has previously been widely employed to induce controlled unidirectional out-of-plane lattice expansion in oxides and semiconductors\cite{guo2015straindoping,brahlek2023emergent,herklotz2025polarization,kim2024suppression,herklotz2025modulating}. 
By performing Helium implantation through a lithographically defined mask, we generate large transverse strain gradients of the order $10^{6}$~m$^{-1}$ in epitaxial GdAuGe films grown on graphene/Ge(111). This approach provides a wafer-scale and highly reproducible route to impose locally defined lattice expansion while avoiding the fragility of freestanding membranes and the uncontrolled strain fields associated with relaxation-driven gradients. We observe a strain gradient driven paramagnetic to ferromagnetic phase transition in contrast with homogeneous strains which do not induce a phase transition. Using nanobeam diffraction and magnetic force microscopy we are able to spatially resolve magnetic response consistent with linear flexomagnetic coupling between magnetization and strain gradient. %in agreement with our phenomenological theory. 
Our method provides a crucial first step towards controlling the spatial distribution and type of strain gradient which could help benchmark emerging theories of flexomagnetism.

\section*{Homogeneous strain doping}

\begin{figure}[t]
    \centering
    \includegraphics[width=0.5\textwidth]{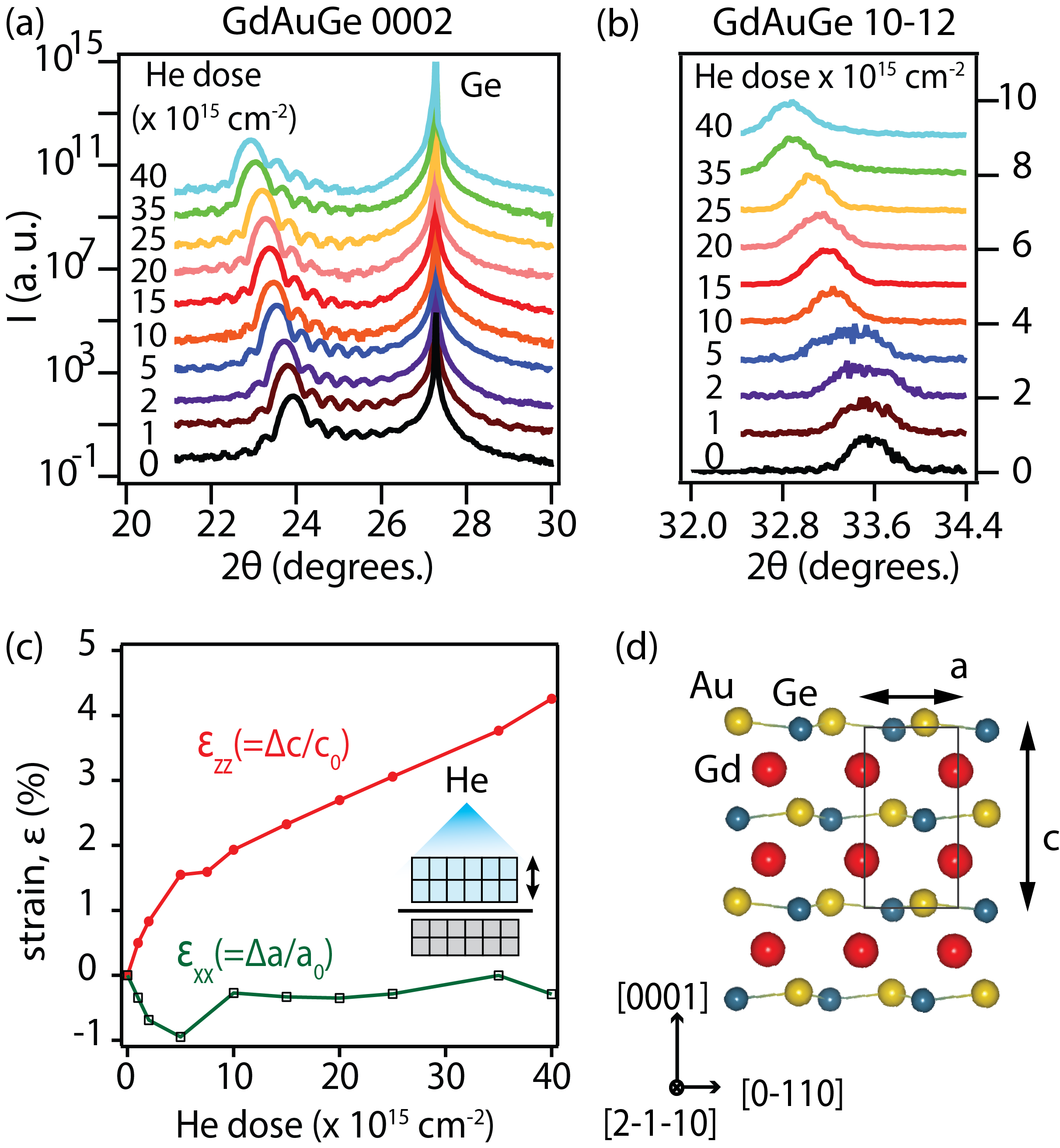}
    \caption{\textbf{Homogeneous strain doping.} (a) X-ray diffraction $\theta-2\theta$ scans about the (0002) GdAuGe reflection fo He implanted GdAuGe films on graphene/Ge (111) substrates. (b)~Off-axis 10$\bar{1}$2 reflection. (c)~Variation of the out-of-plane (c-axis) and in-plane (a-axis) lattice constants with helium dose, with the percentage values indicating the relative change compared to the undosed state. (d)~The unit cell of GdAuGe showing the unit cell constants.}
    \label{strain}
\end{figure}

We first demonstrate homogeneous strain doping of GdAuGe by introducing helium at energy 5 kV with varying ion fluences (doses). The GdAuGe films are grown by molecular beam epitaxy (MBE) on graphene/Ge (111) \cite{laduca2024cold}. Here, low energy Helium implantation offers a proven way to achieve large unidirectional out of plane lattice expansions via chemical pressure or \textit{strain doping} with exceptional spatial and depth control\cite{guo2015straindoping,herklotz2025polarization}. GdAuGe is a topological nodal line semimetal which has an antiferromagnetic (AFM) ground state with many close lying degenerate AFM spin configurations and metamagnetic transitions \cite{ram2023multiple, kurumaji2024metamagnetism}. It has a non-centrosymmetric layered hexagonal structure (space group: $P6_3mc$), where Au and Ge atoms form a wurtzite-type network, and Gd atoms form a triangular lattice in the $ab$ plane (Fig \ref{strain}d). The inherent magnetic interactions in this system are dominated by long-range Ruderman-Kittel-Kasuya-Yosida (RRKY) interactions; hence the ground state magnetic properties are susceptible to changes in the bond distances. 

Figure \ref{strain}(a) presents x-ray diffraction measurements around the GdAuGe (0002) out-of-plane reflections, for varying He doses from 1$\times10^{15}$ ions/cm$^2$ to 40$\times10^{15}$ ions/cm$^2$. Persistent Kiessig fringes confirm homogeneity and interface quality of the GdAuGe, below thresholds for detectable He-induced damage. The (0002) film reflection shifts to lower 2$\theta$ angles with increasing Helium doses, corresponding to c axis expansion. Changes in the in-plane lattice parameter ($a$) were also measured through off-axis scans of 10$\bar{1}$2 reflection as shown in Fig.\ref{strain}(b). The derived change in the unit cell parameters $\Delta$c/c$_0$ and $\Delta$a/a$_0$ as a function of helium doses are plotted in Fig.\ref{strain}(c). We see a clear linear increase in the $c$ parameter indicating a large out of plane strain, $\varepsilon_{zz}$ of 4.3\% with a dose of 40$\times10^{15}$/cm$^2$, whereas $a$ remains fairly constant with the dosing parameter indicating only a minor shift of the in-plane strain, $\varepsilon_{xx}$. Comparative reciprocal space maps of 10$\bar{1}$6 reflection of the pristine system and two heavily helium implanted systems are shown in the supplement Fig.\ref{RSM_uniform}(a) to (c)). We observe large out-of-plane strains up to $\varepsilon_{zz} = 4.2\%$ compared to the minor in-plane strains of $\varepsilon_{xx} = -0.2$ to $-0.3\%$
%a compressive in-plane strain around 0.2-0.3\% in the two systems, while the only measurable change is along the $Q_z$ axis, showing a clear expansion along the c-axis. 

SQUID measurements reveal that this uniform uniaxial strain, even as large as 4.3\%, preserves the antiferromagnetic ground state, with only small shifts in $T_N$ (Supplemental Fig.~\ref{mag_uniform}). Density functional theory (DFT) simulations corroborate this insensitivity: the antiferromagnetic configuration remains as the ground state under uniaxial strains out-of the plane up to 10\% and under biaxial strains in the plane of the order $\pm$3–4\%. [see supplementary figure \ref{DFT_strain} for more details].

\begin{figure}[t]
    \centering
    \includegraphics[width=0.45\textwidth]{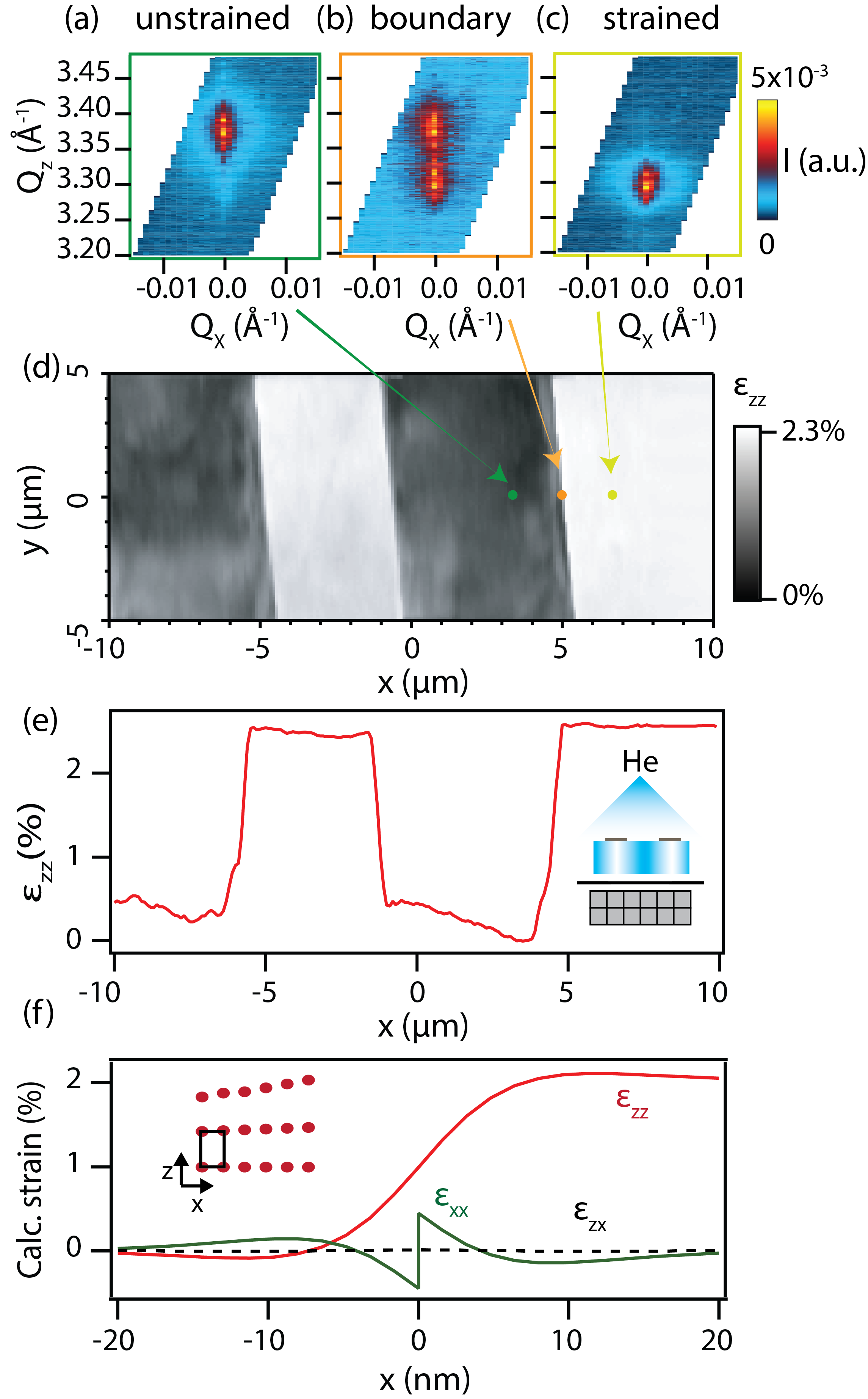}
    \caption{\textbf{Patterning of transverse strain gradients.} (a)-(c) Representative 2D reciprocal space maps (RSM) using hard-xray-nanobeam of 40 nm effective beam width at uniformly strained, boundary and unstrained positions (denoted by markers). The vertical axis $Q_z$ follows the reciprocal lattice vector of (0004) reflection. (d) Spatial strain map constructed by analyzing the 2D RSMs recorded at each (x,y) sample position over an area of 20 $\mu$m $\times$10 $\mu$m with steps of 100 nm. The color-scale on the left denotes the out of plane strain percentage. Details of strain calculation is described in supplementary text. (e)~Experimental strain profile extracted from the strain-map shown in (d). (f)~Theoretical strain relaxation profile over a single strain-unstrained phase boundary, calculated using phase-field simulations. The strain components shown are out of plane ($\varepsilon_{zz}$), in plane  ($\varepsilon_{xx}$) and shear ($\varepsilon_{zx}$) strains.
    } 
    %[\textcolor{red}{strain\% on the strain map needs to be updated and change to color map-Hanfei.}]}
    \label{pattern}
\end{figure}

\begin{figure*}[t]
    \centering
    \includegraphics[width=0.8\textwidth]{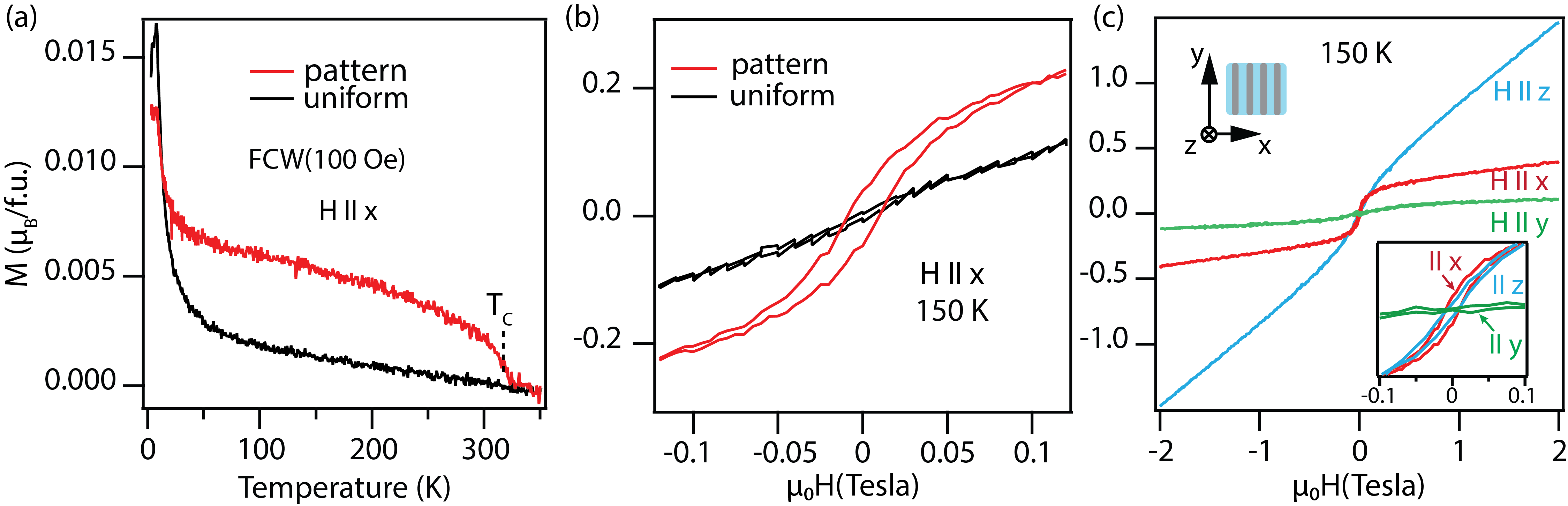}
    \caption{\textbf{Strain gradient-induced ferro/ferrimagnetism.} (a)~Temperature dependence of magnetization under 100 Oe magnetic field oriented inplane. The strain patterned film has alternative stripes with $\varepsilon_{zz}=0\%$ and  $\varepsilon_{zz}=2\%$. The uniform sample is homogeneously doped to $\varepsilon_{zz}=2\%$. (b)~Isothermal magnetization measured at 150 K with magnetic field variation upto $\pm$ 0.1 Tesla. The strain-patterned sample shows a non-linear ferromagnetic hysteresis loop with remanence of 0.06 $\mu_B$ and coercivity of $~$100 Oe, whereas the uniformly strained sample shows a linear response with negligible remanence and coercivity. Field orientation is inplane along x. (c)~Field-direction dependence of isothermal magnetization measured at 150 K in the strain-patterned sample. A nonlinear hysteresis loop which saturates with field, appears only when applied field is along the direction of strain propagation (x). Inset shows the zoomed-in view of the hysteresis in all three direction.}
    \label{flexo}
\end{figure*}

\section*{Strain gradient-induced ferromagnetism}

%We next create transverse strain gradients $\partial \varepsilon_{zz}/\partial x$ by implanting He through a laterally defined photoresist mask.
%Our mask consists of a 10 $\mu$m period mask with alternating strained and unstrained regions. The persistence of sharp Kiessig fringes in X-ray diffraction (XRD) scans (supplement Fig.\ref{RSM_pattern}(a)) confirms that the pattern film interfaces with the substrate remain atomically smooth and that overall crystallinity is preserved. Unlike existing strain-gradient tuning methods like bending and twisting, which are unable to control a single crystal axis, here we have independent and continuous control of a single strain gradient component. In this case, the resulting strain gradient is predominantly transverse ($\partial \varepsilon_{zz}/\partial x$), since the film's in-plane lattice remains largely unaffected by helium implantation. 
We next create transverse strain gradients $\partial \varepsilon_{zz}/\partial x$ by implanting helium through a laterally defined photoresist mask. Our mask consists of a 10 $\mu$m period stripes with alternating helium-exposed (strained) and unexposed (unstrained) regions. The persistence of sharp Kiessig fringes in laboratory X-ray diffraction (XRD) scans (supplement Fig.\ref{XRD_pattern}(a)) confirms that the patterned films retain atomically smooth interfaces and high crystalline quality following He implantation.

Unlike strain-gradient tuning approaches like bending and twisting, which inherently generate multiple strain-gradient components, our lithographically defined implantation geometry enables independent and continuous control of a single dominant strain gradient component, while minimizing competing gradients. In the present case, the resulting strain gradient is predominantly transverse ($\partial \varepsilon_{zz}/\partial x$), as helium implantation induces a large out-of-plane lattice expansion that varies in the plane, while leaving the in-plane lattice parameters largely unchanged.

We spatially map the strain gradient using nanobeam diffraction at the hard X-ray nanoprobe (HXN) beamline of the National Synchrotron Light Source II, employing an effective beam diameter of 40 nm. A 20~$\times$~10~$\mu$m$^2$ area was raster-scanned with a 100~nm step size in both lateral directions ($x$ and $y$) while recording the GdAuGe (0004) Bragg condition. Representative RSMs acquired at positions corresponding to the unstrained, strained, and boundary are shown in Fig.\ref{pattern}(a)-(c).

At each $(x,y)$ position the local out of plane effective strain $\varepsilon_{zz}$ was calculated from a single Gaussian fitting of the corresponding RSM peak, yielding the spatial strain map shown in Fig.\ref{pattern} (d). The line profile in Fig.~\ref{pattern}(e) demonstrates the lateral continuity of the patterned strain along the ($x$) direction. We observe a contribution from both the strained and unstrained diffraction conditions at the boundary of helium-implanted and pristine regions [see Fig.\ref{pattern} (b)]. Consequently, $\varepsilon_{zz}$ at each boundary is obtained from the weighted average of this bimodal diffraction signal, leading to an experimentally inferred effective strain gradient $\partial \varepsilon_{zz}/\partial x$ that is laterally broadened over $\approx$ 750~nm. 
%\textcolor{red}{[mark this broadening on the figure? Is it defined as the length over which there are two clear Gaussians in supp fig S5b?]}
A more detailed analysis of strain gradient broadening across the boundary, accounting for bimodal diffraction signal using two-Gaussian fitting, is provided in the supplementary information (Fig.~\ref{nanobeam-SM}).

%At each $(x,y)$ positions the out of plane strain $\varepsilon_{zz}$ was calculated from the center of mass of the corresponding RSM peak, yielding the spatial strain map shown in Fig.\ref{pattern} (d). The line profile in Fig.~\ref{pattern}(e) demonstrates the continuity of the patterned strain along the horizontal ($x$) direction. We note that we observe diffraction contribution from both unstrained and strained (0004) reflection at the interfaces [see Fig.\ref{pattern}]. Consequently, $\varepsilon_{zz}$ at each boundary is determined from the center of mass of this bimodal diffraction signal. As a result, the inferred strain gradient $\partial \varepsilon_{zz}/\partial x$ has a relaxation over a lateral distance of $\approx$ 100~nm. 

To further explore the strain relaxation, we employed mechanical modeling of the elastic relaxation. Calculations based on phase field simulation\cite{chen2022from} were performed considering a 2D GdAuGe system with alternating pristine and strain states in the $x-z$ plane to understand the behavior of each contributing component of the strain profile. As shown in Fig.\ref{pattern} (c), the out of plane strain ($\varepsilon_{zz}$) relaxes over the length scale of approximately 20 nm as one approaches equilibrium, which is too sharp to be measured using an experimental probe. Contributions of the in-plane strain component ($\varepsilon_{xx}$) and the shear component ($\varepsilon_{zx}$) are also plotted in the same graph, clearly indicating their negligible effects on the overall strain gradient in this system. %out of plane 2\%, inplane 0.2%, shear 0.009\% 
We estimate an upper bound of the local strain gradient here to be 2\% per 20 nm, which is equivalent to 1$\times10^6$/meter. From the experimentally observed strain boundary, we estimate a lower bound to be $2.7\times10^4$/meter corresponding to a 2\% strain variation over a 750 nm length scale.

%$\textcolor{red}{[compare this to a lower bound estimate based on the experimental broadening? 2 percent over a length scale of 1 micron? Also is gradient in units of percent per meter? Or fraction per meter where 100 percent equals 1?]}
The strain-gradient magnitudes we achieve place our system directly in the regime where they compare favorably with other platforms used to engineer gradient-driven responses. Bent crystalline membranes such as ultrathin SrTiO$_3$ or transition-metal dichalcogenides typically generate curvature-induced gradients in the $10^5-10^6$/meter range, but only under mechanical deformation and with limited spatial patternability\cite{tang2025flexomagnetism}. Flexoelectric oxide heterostructures also routinely exploit strain gradients of $10^5-10^7$/meter to induce large polarization responses\cite{shen2024flexoelec}, demonstrating that our values lie at the upper end of established functional gradient-engineered systems.

\begin{figure*}[t]
    \centering
    \includegraphics[width=0.8\textwidth]{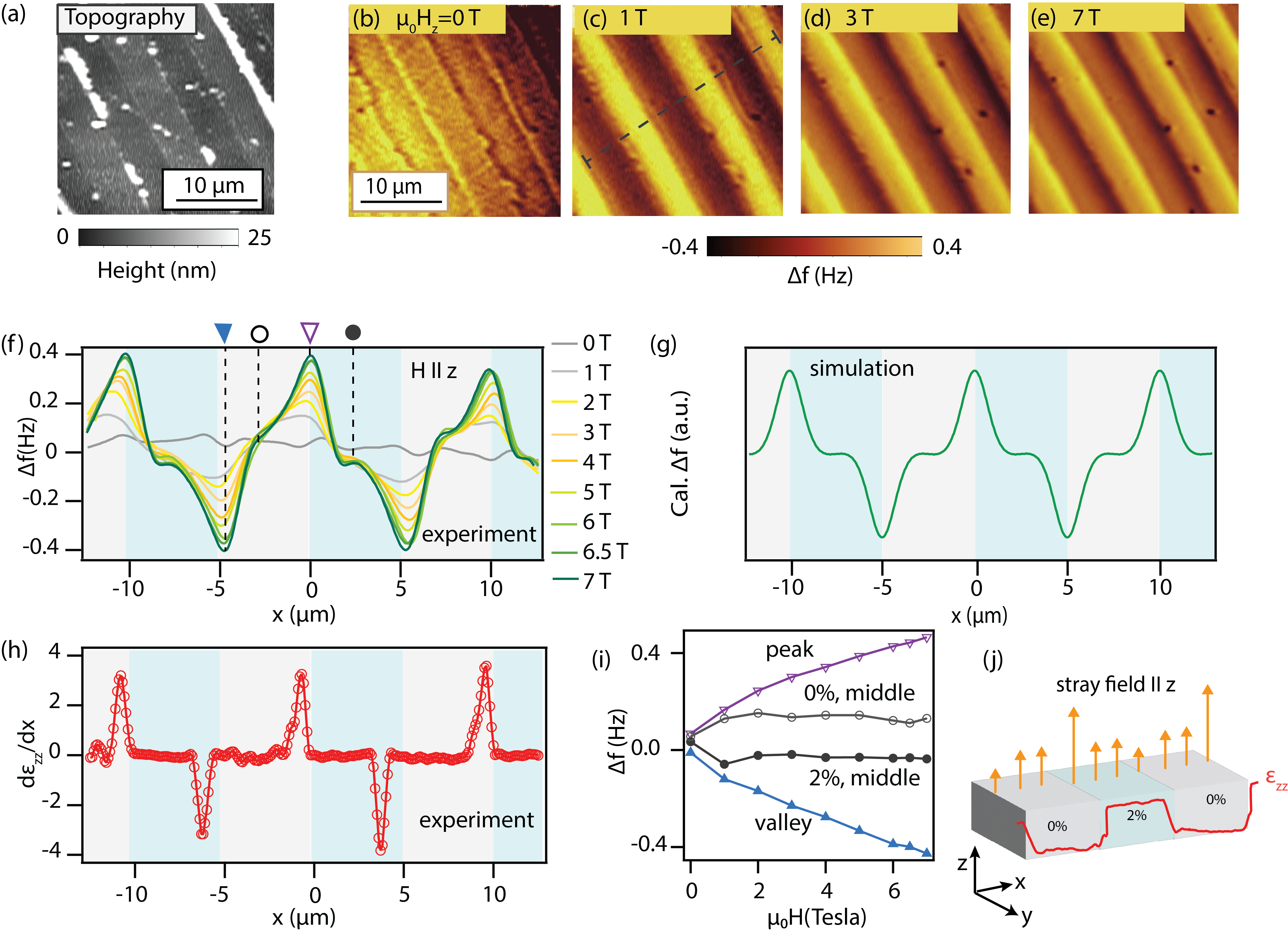}
    \caption{\textbf{Magnetic force microscopy (MFM) of the strain-patterned GdAuGe.} (a)~Atomic force microscopy topographic image of the sample surface, higher steps (brighter) are helium-exposed or strained regions. The randomly distributed white spots and stripes are residual photoresist. (b)-(e)~Representative MFM images taken at 20 K under 0, 1, 3, and 7 Tesla out of plane magnetic fields $H_z$. (f)~Line profiles of the MFM signal ($\Delta f$) taken at the position of the dotted line in (c). The shaded background marks the strained and unstrained regions.(g)~Micromagnetic simulations of MFM signal $\Delta f$ considering a stray field distributions shown as schematic in (j). %\textcolor{red}{with tip-sample separation XX 1 um?}. 
    (h)~Experimental strain-gradient derived from nanobeam diffraction profile in Fig.\ref{pattern}(b). (i)~Magnetic field dependence of the $\Delta f$ values at locations marked in (f). %pristine to strain boundary (peaks), middle of 2\% strained region, middle of pristine region, and at strain to pristine boundary (valley). 
    (j) Schematic describing the stray field in the strain patterned GdAuGe film.
    %\textcolor{red}{b: update labels to say ``(b) $\mu_0 H_z = 0$T'' to emphasize out of plane applied field}
    }
    \label{MFM}
\end{figure*}

%\textcolor{red}{[lead with: we find that strain gradients induce a para to ferro/ferrimagnetic transition... and compare data by strain patterning to our previous demonstration for bending]}
%\textcolor{blue}{question may come up about the large difference between the T$_C$ and T$_N$.}

We find that the patterned strain gradients in GdAuGe drive a magnetic phase transition from paramagnetic to ferromagnetic like state near room temperature.  
%An antiferromagnetic-to-ferromagnetic transition has previously been observed in freestanding GdAuGe membranes subjected to bending-induced strain gradients of order $3.6\times10^{5}$~m$^{-1}$ at wrinkle crests and troughs\cite{laduca2024cold}. In contrast to bending geometries, which inherently generated mixed longitudinal and transverse strain-gradient components, the patterned films studied here predominantly realize a single transverse component, $\partial \varepsilon_{zz}/\partial x$, allowing a more direct connection between a single tensor element of the flexomagnetic response and the magnetic ground state.
Figure~\ref{flexo}(a) shows the temperature dependence of the field-cooled warming (FCW) magnetization, measured at 100~Oe with the field applied parallel to the propagation direction of the strain gradient (x-axis). 
%For comparison, the $M(T)$ curve of a uniformly strained (2\%) control sample is also shown. 
Both patterned and uniformly strained (nominally 2\%) exhibit a peak near $T_N \approx 7$~K, reflecting the antiferromagnetic ordering intrinsic to uniformly strained GdAuGe. 
In the patterned sample, a pronounced ferromagnetic-like increase in $M(T)$ emerges immediately above $T_N$ and persists up to a temperature, $T_C$ of approximately 318~K, whereas the uniformly strained control reverts directly to a paramagnetic state above $T_N$. A similar paramagnetic-to-ferromagnetic transition has previously been observed in rippled GdAuGe \cite{laduca2024cold} and GdPtSb \cite{du2021epitaxy} membranes with mixed longitudinal and transverse strain gradients. An advantage of the new patterned He implant geometry is that we are able to isolate the effects of a transverse gradient.

The temperature difference between the two magnetic transitions in the strain patterned sample is substantial ($T_N \approx 7$ K, $T_C \approx 318$ K). However, such behavior is not uncommon in rare-earth intermetallics whose magnetism is governed by Ruderman–Kittel–Kasuya–Yosida (RKKY) exchange\cite{tseng2007effect, ruderman1954indirect, kasuya1956progress,yosida1957magnetic}. In these systems, the ordering temperature is highly sensitive to the $4f-5d$ hybridization and conduction-electron polarization, making the magnetic ground state exceptionally responsive to changes in lattice geometry, local bonding, and exchange pathways. As a result, relatively modest structural perturbations can drive shifts of hundreds of kelvin in transition temperatures through changes in the sign and magnitude of the RKKY interaction\cite{tseng2007effect}. In GdAuGe, we find that the uniform out-of-plane strain does not change the antiferromagnetic ground state configuration, indicating that homogeneous lattice expansion alone is insufficient to alter the exchange interactions here. By contrast, the introduction of a transverse strain gradient produces spatially varying interatomic distances and locally breaks the lattice symmetry at the patterned boundaries. These gradient-induced distortions modify the local $4f-5d$ hybridization and conduction-electron polarization, thus reshaping indirect exchange interactions and stabilizing the ferromagnetic order at temperatures far above $T_N$.

The strain-patterned samples exhibits a pronounced nonlinearity and hysteresis in the isothermal magnetization ($M(H)$) which are hallmarks of a ferromagnetic behavior (Fig.\ref{flexo}(b)). The magnetization value reaches $\approx$ 0.2 $\mu_B$/f.u. at 0.1 Tesla with coercive field ($H_C$) of $\sim$ 100 Oe, measured at 150 K. 
The magnetization value of 0.2 $\mu_B$/f.u. represents a moment averaged over the entire film. Because only a fraction of the film volume in the strain-gradient regions contributes to the ferromagnetic response, the local magnetization within these regions is therefore expected to be substantially larger.
In contrast, the uniformly strained control sample shows a monotonic linear increase in magnetization values with no sign of saturation and zero coercivity ($H_C$) typical of a paramagnet (Fig.\ref{flexo}(b)). 
Low-energy He implantation was employed to minimize implantation-induced damage, and uniformly dosed control samples do not exhibit ferromagnetic signatures, indicating that the observed magnetic response is linked to strain gradients rather than being dominated by implantation-induced defects.
%\textcolor{red}{[The strong in-plane anisotropy also points towards a strong flexomagnetic effect, as opposed to an effect from uniform strain?]}

The patterned sample also displays a clear magnetic anisotropy with applied magnetic field orientation (Fig.\ref{flexo}(c)). The $M(H)$ along the strain gradient ($x$ direction) shows a clear saturation compared to the $M(H)$ with a field perpendicular to the strain gradient ($y$ direction) and the field out of the plane ($z$ direction). 
$H_C$ (inset of Fig.~\ref{flexo}c) is pronounced along the strain-gradient direction and in the out-of-plane geometry, while the response perpendicular to the gradient remains paramagnetic-like. These results together demonstrate that the transverse strain gradient induces a magnetic phase transition with pronounced hysteresis and field dependent magnetic anisotropy.
%generates from a may not be readily described by a simple linearly coupled $M$ and $\frac{\partial \varepsilon_{jk}}{\partial x_l}$ relation in Eq.\ref{eq1:flexo}, as it does not account for the gradient-induced anisotropy and domain pinning that produce the finite, hysteretic $H_c$ and a saturating $M_x$ for $H\parallel x$.

\section*{Local magnetic mapping}

Magnetic force microscopy (MFM) measurements with out of plane field $H_z$ confirm that the flexomagnetic response is localized to the boundaries between strain and unstrained regions, where strain gradients are maximized. 
%\textcolor{red}{Here we use the atomic force microscopy (AFM) topography to establish spacial registry between the MFM-based magnetic mapping and the nanobeam diffraction mapping of strain.}
We observe a one-to-one correspondence between the periodic atomic force microscopy (AFM) topographic features due to masked helium implantation [Fig.~\ref{MFM}(a)] and the field-dependent MFM signal shown in  Figs.~\ref{MFM}(b–e). Here the measured frequency shift $\Delta f$ is proportional to the out-of-plane magnetic force gradient\cite{Feng2022}. The line profiles in Fig.~\ref{MFM}(f) reveal that MFM signal $\Delta f(x)$ at high field (e.g.~7~Tesla) display alternating peaks and valleys at successive boundaries, where the magnitude of $\partial \varepsilon_{zz} / \partial x$ is largest and flips sign. %This alternating behavior qualitatively matches the transverse strain gradient, which is consistent with the flexomagnetic mechanism described in Eq.\ref{eq1:flexo} (Discussion). 
%Combined topography and MFM at 20~K [Figs.~\ref{MFM}(a–e)] reveal a spatially periodic modulation of the MFM frequency-shift signal $\Delta f$ locked to the strain gradient pattern. $\Delta f$ is proportional to the local magnetic stray field behavior perpendicular to the sample surface here. In Fig.\ref{MFM}(f) and (g), $\Delta f(x)$ reverses sign at every strain boundary and reproduces the odd spatial symmetry of the strain gradient, consistent with a flexomagnetic origin described in Eq.\ref{eq1:flexo}. 
%This behavior differs from typical MFM responses observed at topographic step boundaries or antiferromagnetic domain walls under external magnetic fields \cite{park2009pancake, sass2020magnetic},  where nominally equivalent boundaries are expected to produce similar magnetic responses. %\textcolor{blue}{[can we make this statment more specific?... do we mean that AFM domains should not produce alternativn peaks and valleys? instead they should have the same sign, i.e. all peaks or all valleys? Also try to make a more direct comparison ruling out piezomangetism here...]}
This behavior differs from typical MFM responses observed at topographic step boundaries under external magnetic fields \cite{park2009pancake}, where $\Delta f$ consists of antisymmetric up-down features localized at each step edge, which differs from the peak and valley response we observe here. It also differs from antiferromagnetic domain walls\cite{sass2020magnetic} under external magnetic field, which produce a single localized extrema at each domain wall without a systematic alternation of peaks and valleys at neighboring walls.
%By contrast, the GdAuGe patterned films show periodic modulation, strict synchrony with the strain pattern, and an increasing $\Delta f(H)$ response persisting at 7~T [Fig.~\ref{MFM}(f)], which is incompatible with canted-AFM domain-wall contrast between two AFM configurations or magnetostatic self-organization generated by small uncompensated moments over a narrow domain wall. In such cases, the MFM signal collapses once a moderate out-of-plane field removes canting or polarizes the domains. 

Interestingly, the peaks and valleys of the MFM signal are significantly broader ($\Delta x_{MFM}\sim$ few microns) than the measured peaks and valleys of the strain gradient $\partial \varepsilon_{zz}/\partial x$ ($\Delta x_{straingadient}\sim$ 750 nm). The broadness of this experimental MFM signal might be caused by an enhanced effective tip-sample distance. The exact mechanism is unclear at this moment. 

To further validate the flexomagnetic origin, we performed micromagnetic simulations of the MFM response [Fig.~\ref{MFM}(g)] based on the stray-field distribution illustrated in Fig.~\ref{MFM}(j). Consistent with experimental observations, the simulated profile displays alternating behavior at the locations of strain gradient variations, reproducing the key features of the data. The simulation also captures the spatial extent of the measured signals. The observed broadening of the MFM signal at boundaries is consistently reproduced in the simulation with the tip-sample distance $\sim1~\mu$m. Details of the simulation are provided in the methods section H. 

Figure~\ref{MFM}(i) shows the field dependence of the MFM signal at selected locations (marked in Fig.~\ref{MFM}(f)), showing that the magnitude of the signal at the strain boundaries increases linearly with the applied magnetic field for both polarities. In contrast, uniform regions, whether strained (nominally 2$\%$) or unstrained (0$\%$), show small response that saturates at a low field ($\sim1$~Tesla). This behavior reflects a susceptibility response that is spatially localized to the strain boundaries, rather than being uniformly present throughout the strained regions as in bulk piezomagnetic effects. Note that the zero level of the MFM signal serves as a reference baseline, and its sign reflects relative changes of local magnetic susceptibility.
Taken together, these observations indicate that a linear, odd-in-strain-gradient magnetic contribution manifests primarily as a local modulation of the out-of-plane magnetization, which is directly resolved by MFM and averages to a nearly zero net moment in macroscopic $M(H_z)$ measurements.

\section*{Discussion}

We perform a symmetry analysis to rationalize the alternating peak and valley MFM contrast.
Based on a phenomenological theorem\cite{Sai2018flexomagnetic,lukashev2010flexomagnetic}, the magnetic response in the presence of strain $\varepsilon$ and the strain gradient $\partial \varepsilon_{jk} / \partial x_l$ is:
\begin{equation}
    M_{i} = \chi_{ij}H_j + c_{{ijkl}} \varepsilon_{{ij}} \varepsilon_{{kl}} +d_{ijk} \varepsilon_{jk}+\mu_{ijkl} \frac{\partial \varepsilon_{jk}}{\partial x_l}
    \label{eq1:flexo}
\end{equation}
where, $\chi_{ij}H_j$ is the intrinsic magnetic response of the material and $\chi_{ij}$ is the magnetic susceptibility. $c_{ijkl}$ and $d_{ijk}$ are magneto-elastic and piezomagnetic tensors, respectively. $\mu_{ijkl}$ represents the flexomagnetic tensor, which couples $M$ linearly with the strain gradient and exhibits symmetry properties distinct from those of $c_{ijkl}$ and $d_{ijk}$\cite{eliseev2011linear}. However, the change in thermodynamic potential and consequently the magneto-mechanical response of a material due to strain gradient is complex at the nano- and micro-scale\cite{Sai2018flexomagnetic}. Such multi-field studies require consistent boundary conditions that depend on both the symmetry and the control of specific strain gradient field of a material.

%To rationalize the alternating MFM contrast and the linear field dependence, we consider the leading-order strain-gradient dependence of the magnetic susceptibility.
To connect with MFM experiments, we consider the leading-order strain-gradient dependence of the magnetic susceptibility.
\begin{equation}
    \label{eq2:flexopara}
\chi_{ij}(\mathbf{r})=\chi_{ij}^{(0)} + \gamma_{ijklm}\,\frac{\partial\varepsilon_{lm}}{\partial x_k}(\mathbf{r}).
\end{equation}
Here,  $\chi_{ij}^{(0)}$ is the uniform magnetic susceptibility of the material, i.e. how magnetization responds to the magnetic field in the absence of any strain gradient and $\gamma_{ijklm}$, which is a fifth rank tensor, describes how the paramagnetic susceptibility changes in response to a strain gradient. The flexoparamagnetic term is odd in the strain gradient and therefore changes sign when the gradient is reversed. Being a fifth rank tensor, $\gamma_{ijklm}$ should also be nonzero in a larger number of point groups, i.e. it should be constrained more weakly by symmetry [see Supplementary Note B for more details].
Eq.\ref{eq2:flexopara} yields the local magnetization
\begin{equation}
M_i(\mathbf{r}) = \left[\,\chi_{ij}^{(0)} + \gamma_{ijklm}\,\frac{\partial\varepsilon_{lm}}{\partial x_k}(\mathbf{r})\,\right]H_j.
\end{equation}

For hexagonal $6mm$ symmetry of paramagnetic GdAuGe with %an in-plane shear strain gradient $\frac{\partial\varepsilon_{xz}}{\partial x}$ and 
field along $z$, the linear susceptibility response reduces to: %this reduces to
\begin{equation}
    M_z(x)=\left[\,\chi_{zz} + \gamma_{zzxzx}\,\frac{\partial\varepsilon_{zx}}{\partial x}\right]H_z.
\end{equation}
where only a shear strain gradient $\partial\varepsilon_{zx}/\partial x$ is allowed to produce a linear change in the out of plane magnetic susceptibility $\chi_{zz}$. Normal strain gradient is symmetry forbidden for a linear response in $\chi_{zz}$. Yet our experiments and phase field modeling (Fig.\ref{pattern}) suggest that shear strain gradients are small compared to the normal transverse strain gradient $\partial\varepsilon_{zz}/\partial x$. Therefore the observed correlation of the magnetic signal with transverse strain gradient cannot be explain within linear paramagnetic susceptibility theory alone. This implies a symmetry breaking may be the cause for the strong apparent dependence on normal strain gradient $\partial\varepsilon_{zz}/\partial x$. Such a symmetry breaking is plausible, given the paramagnetic to ferromagnetic phase transition that we observe by SQUID magnetometry (Fig. \ref{flexo}).

The strain-gradient landscape also introduces additional internal anisotropy and pinning fields that are not captured by the linear flexoparamagnetic coupling term in Eq.~\ref{eq2:flexopara}. These nonlinear contributions give rise to hysteresis in both $M(H_z)$ and $M(H_x)$, but with markedly different field responses. While the present work establishes strain-gradient-driven magnetic responses, a detailed microscopic description of the underlying flexomagnetic mechanisms is beyond its scope. A complete quantitative determination of flexomagnetic effects and field dependence will require probes with calibrated stray-field sensitivity such as NV-center magnetometry or nanoscale scanning SQUID or nanoprobe xray magnetic circular dichroism. Nonetheless, the spatial correlation between the strain gradient and the alternating MFM signal, combined with the emergence of high-temperature ferromagnetism in global measurements, establishes patterned helium implantation as a robust and deterministic route to engineer flexomagnetic responses in thin films of quantum materials.

In summary, we have demonstrated that lateral strain gradients engineered through patterned helium ion implantation provide a robust and deterministic route to control magnetism in thin films of the antiferromagnetic semimetal GdAuGe. By isolating a single, predominantly transverse strain-gradient tensor component, $\partial\varepsilon_{zz}/\partial x$, we uncover a strain-gradient–induced magnetic phase transition that drives the native low-temperature antiferromagnet to a ferromagnetic-like state with a transition temperature extending to room temperature. The strain-gradient landscape generates strong internal anisotropy and pinning, stabilizing ferromagnetic-like domains with a preferred easy axis along the gradient direction. By combining highly localized spatial probes as Nanobeam x-ray diffraction and magnetic force microscopy we resolve that the local magnetization is defined by the sign-alternating strain gradient with a periodicity inherited from the lithographically designed strain pattern. Our results establish patterned strain gradients as an experimentally accessible platform for tuning magnetic interactions beyond what is possible with uniform strain, chemical substitution, or external fields. More broadly, they demonstrate how flexomagnetic design principles can be integrated into quantum materials to create spatially programmable magnetic textures, gradient-defined anisotropies, and strain-controlled phase transitions.

%Therefore, in future, a fully quantitative determination of the corresponding tensor element $\gamma_{zzzxz}$ will require a probe capable of mapping the absolute stray field or local magnetization with calibrated sensitivity, for example, nanoscale scanning SQUID or NV-center magnetometry.

\section{Methods}
\subsection{Sample synthesis and helium implantation}
The GdAuGe single crystalline thin films were grown on graphene/Germanium (111) substrate by a cold seeded epitaxy approach using molecular beam epitaxy. The detailed growth mechanism is set forth in ref\cite{laduca2024cold}. The thickness of the films used for helium (He) implantation is nominally 20 nm thick. The samples were introduced to different He doses using a SPECS IQE 11/35 ion source at Oak-Ridge National Laboratory. During implantation, the acceleration voltage was fixed at 5 keV, with the beam flux calibrated to 1.56$\times$10$^{12}$ He/cm$^2$ per second using a Faraday cup. Helium is particularly advantageous for controlled ion implantation because its stopping power within the lattice is dominated by non-nuclear interactions, thereby minimizing structural damage to the film\cite{guo2015straindoping}. In addition, the chemical inertness of helium ensures that it does not introduce extra charge carriers and limits the formation of defects\cite{livengood2009subsurface}. An estimation of the Helium doses to be implanted were calculated by Monte Carlo simulations of the stopping range and ion motion, considering a low energy beam and the thickness of the GdAuGe films. Further the experimental doses were carefully calibrated, reducing defect formation and eliminating surface sputtering effects. The lithography patterning of the samples was performed at Oak Ridge National Laboratory with a hard mask of the 5 $\mu$m stripe pattern using standard photolithography. Before patterning, the in-plane orientation of the sample was confirmed by XRD so that the stripe patterns may be oriented parallel to the crystallographic axis $21\bar{1}0$. %perpendicular to the Ge[0$\bar{1}$1] axis. 
To clean the samples for photolithography, samples were treated with an acetone, IPA and DI water wash to clean the surface, followed by an immediate nitrogen blow dry upon removal from the DI water. After development, the exposed stripe regions were implanted with He ions.  %The sample's $21\bar{1}0$ crystallographic axis was aligned to be parallel to the stripes. 

\subsection{Structural characterization}
Lab based XRD measurements were performed with Cu K$\alpha$ radiation and a four circle geometry using a Malvern PANalytical Empyrean diffractometer at University of Wisconsin Madison and PANalytical MRD at Oak Ridge National Laboratory.
X-ray diffraction (XRD) peak positions were obtained by fitting the measured intensity profiles to extract the Bragg angle $2\theta$. The corresponding interplanar spacing was calculated using Bragg’s law (first order, $n=1$):
\begin{equation}
d_{hkil} = \frac{\lambda}{2\sin\theta},
\label{eq:bragg}
\end{equation}
where $\theta = (2\theta)/2$ and $\lambda$ is the X-ray wavelength.

For the hexagonal lattice of GdAuGe, the lattice parameters are related to the interplanar spacing by
\begin{equation}
\frac{1}{d_{hkil}^{2}} =
\frac{4}{3}\frac{h^{2}+hk+k^{2}}{a^{2}} +
\frac{l^{2}}{c^{2}},
\label{eq:hex_metric}
\end{equation}
using the Miller--Bravais indexing scheme ($i=-(h+k)$). The in-plane lattice parameter $a$ and out-of-plane lattice parameter $c$ were extracted by fitting Eq.~\ref{eq:hex_metric} to the experimentally determined $d_{hkil}$ values.

For purely out-of-plane reflections $(000l)$, the expression simplifies to $d_{000l}=c/l$, yielding
\begin{equation}
c = \frac{l\lambda}{2\sin\theta}.
\label{eq:c_general}
\end{equation}
In this work, the $(0004)$ reflection was used to determine the out-of-plane lattice parameter, giving
\begin{equation}
c = \frac{2\lambda}{\sin\theta}.
\label{eq:c_0004}
\end{equation}

The in-plane lattice parameter $a$ was extracted from nonzero in-plane reflections $(hkil)$ using
\begin{equation}
a =
\sqrt{
\frac{4}{3}
\frac{h^{2}+hk+k^{2}}
{\displaystyle \frac{1}{d_{hkil}^{2}} - \frac{l^{2}}{c^{2}} }
}.
\label{eq:a_extraction}
\end{equation}

%The uncertainty in the interplanar spacing was obtained by standard error propagation.

\subsection{Magnetic characterization }
Magnetization measurements were performed using a Quantum Design MPMS3 superconducting quantum interference device (SQUID) magnetometer at the department of chemistry, University of Wisconsin Madison, with a sensitivity $<1\times10^{-8}$ emu. Background contributions from the sample holder and substrate were measured separately and subtracted from the raw data. The magnetization values were normalized by sample volume to obtain the magnetization density.

\subsection{Nanobeam hard x-ray diffraction}
Nanobeam X-ray diffraction measurements were performed at the Hard X-ray Nanoprobe beamline (HXN, 3-ID) of the National Synchrotron Light Source II (NSLS-II) at Brookhaven National Laboratory. A monochromatic X-ray beam with an energy of 12~keV was selected using a Si(111) double-crystal monochromator. The beam was focused to a spot size of approximately 40~nm using Fresnel Zone pLate with 30 nm outmost zone width. The sample was mounted on a high-precision piezoelectric scanning stage and positioned at the beam focus.

To enable spatial registration and correction of stage drift during scanning, we lightly dusted the sample surface with copper (Cu) nanoparticles, which served as  positional markers. The sample was raster-scanned in two dimensions with a step size of 100~nm in both horizontal and vertical directions, spanning a total area of 20~$\times$~10~$\mu$m$^2$. The Merlin pixel-array detector (Quantum Detectors) located 400~mm downstream of the sample, with a pixel size of 55~$\mu$m was aligned to detect the prominent 0004 reflection of GdAuGe. Simultaneously, Cu fluorescence signals were collected using a Vortex silicon drift detector (Hitachi) placed orthogonal to the incident beam. The fluorescence signal was used solely to track the position markers and correct stage drift. 

2.1$^\circ$ wide angular scans were performed at 0.1$^\circ$ intervals by rotating the sample through the Bragg condition, enabling the construction of three-dimensional reciprocal space maps (RSMs). %The dwell time at each scan point was 20~ms. 
Nanodiffraction data were processed using Python scripts developed at NSLS-II (\url{https://github.com/NSLS-II-HXN/Nanodiffraction}), which facilitated data loading, alignment, reciprocal space transformation, and strain reconstruction. The relative strain was calculated considering the Gaussian fitting to find the position of the peak in each reciprocal space map. 

Although the nano-focused beam provides high spatial resolution, it also exhibits an angular divergence, which results in elongation along the beam propagation direction in reciprocal space. Depending on the angle of incidence ($\theta$), the beam footprint becomes extended as $x=x_0/sin(\theta)$. For GdAuGe 0004 peak, $\theta \sim$16$^\circ$, which elongates the beam by a factor of 3.6. To minimize the influence of this divergence on the strain analysis, the sample was oriented so that the beam-induced streaking aligned with the horizontal y direction of the sample (sample's length along y, $\sim$1.5 mm). This alignment ensured reliable resolution of strain variations between adjacent stripe domains and the boundaries.

\subsection{Magnetic force microscopy}
MFM and topographic measurements were carried out in a home-built Helium-3 MFM system equipped with a superconducting magnet for applying out-of-plane magnetic fields using commercial piezoresistive cantilevers (spring constant $k\sim3$~N/m; resonance frequency $f_0\sim40$~kHz). The tips were coated with nominally 100~nm cobalt by using magnetron sputtering. MFM images were taken with a non-contact constant-height mode, maintaining a tip-sample separation of 300~nm. The magnetic signal was detected as a shift in the cantilever resonance frequency, which is proportional to the out-of-plane force gradient. The frequency shift was extracted using a phase-locked loop (SPECS). To minimize the contribution from electrostatic forces between the tip and the sample, a DC bias voltage was applied to the sample surface to balance the electrostatic interaction.

\subsection{Density functional theory}
First-principles calculations based on density functional theory (DFT) were performed using the ABINIT software package. The simulations employed the PAW JTH v1.1 pseudopotentials within the local density approximation (LDA) framework. A Monkhorst–Pack k-point grid of $12\times12\times8$ was used to sample the Brillouin zone, with a plane-wave energy cut-off set at 26 Hartree. To account for relativistic effects, spin–orbit coupling was included in all calculations. The DFT+U method was applied to Gd atoms, using a Hubbard U parameter of 10 eV to better describe the localized f-electrons. Structural relaxation was considered complete when the potential residual dropped below 1.0$\times$10$^{-17}$. For evaluating strain effects, the in-plane (out-of-plane) lattice constants were held fixed while allowing relaxation along the out-of-plane (in-plane) direction.

\subsection{Strain distribution simulations}
The equilibrium strain profile is simulated using the elastodynamic solver within the COMSOL Multiphysics software \cite{COMSOL}. The simulation considers a two-dimensional system in the x-z plane (160 nm $\times$ 20 nm) with stress-free boundary conditions on the top ($z = 20$ nm) and bottom ($z = 0$ nm) surfaces. Along the x-axis, the system is divided into two regions, each 80 nm wide, representing the strained and unstrained regions, respectively. The simulations employ rectangular meshes with an average cell size of about 0.5 nm. The model considers the elastodynamic equation Eq.\ref{eq:elastodynamic}:
\begin{equation}\label{eq:elastodynamic}
    \rho \frac{\partial^2 \mathbf{u}}{\partial t^2} = \nabla \cdot (\mathbf{\sigma} + \beta \frac{\partial \mathbf{u}}{\partial t})
\end{equation}
where $\rho$ is the mass density, $\beta$ is the elastic stiffness damping, $\mathbf{u}$ is the mechanical displacement, $\sigma_{ij} = c_{ijkl}(\varepsilon_{kl} - \varepsilon_{kl}^0)$ is the elastic stress tensor. The total strain is related to the mechanical displacement by $\varepsilon_{ij} = \frac{1}{2}(\frac{\partial u_i}{\partial j} + \frac{\partial u_j}{\partial i})$. The eigenstrain $\varepsilon_{ij}^0$ is set to be zero for the unstrained region. For the strained region, $\varepsilon_{zz}^0 = 2\%$ is taken as the only non-zero component. The elastic stiffness constants $c_{ijkl}$ of hexagonal GdAuGe are calculated from the first-principles method: $c_{11} = 164.8468$ GPa, $c_{12} = 81.2855$ GPa, $c_{13} = 61.7714$ GPa, $c_{33} = 105.2717$ GPa, $c_{44} = 53.6384$ GPa, and $c_{66} = 41.7807$ GPa. For the hexagonal GdAuGe, the lab coordinate $(x,y,z)$ refers to the crystal physics coordinate $(x_1,x_2,x_3)=([0\bar{1}10],[2\bar{1}\bar{1}0],[0001])$. The values of the mass density $\rho = 9000$ $kg/m^3$ and the elastic stiffness damping $\beta = 10^{-5}$ $s$ do not affect the final equilibrium state. The spatial distribution of $\mathbf{u}$ reaches its equilibrium when both the first-order and second-order time derivatives of $\mathbf{u}$ are close to zero. The equilibrium distribution of $\mathbf{u}$ is then used to compute the total strain distribution in the 2D plane. The line profile of strain shown in Fig. 3(f) indicates $z = 10$ nm.   

\subsection{MFM signal simulations}
The magnetic force microscopy (MFM) signal is simulated using MuMax3\cite{vansteenkiste2014the}, which computes the tip–sample magnetic interaction from a prescribed magnetization distribution. The simulation is conducted on a one-dimensional sample with periodic boundary conditions along the x-axis (along the strain-pattern) with a cell size of 50 nanometers. Each period has a width of 10 $\mu m$, indicating two neighboring strain-patterned GdAuGe strips with 5 $\mu m$ wide each. The model considers the magnetostatic equation Eq.\ref{eq:magnetostatic}:
\begin{equation}\label{eq:magnetostatic}
\nabla \cdot \mathbf{B} =  \mu_0 \nabla \cdot (\mathbf{H}+\mathbf{M}) = 0
\end{equation}
where $\mu_0$ is the vacuum permittivity, $\mathbf{H}$ is the magnetic field intensity, $\mathbf{M}$ is the out-of-plane saturation magnetization from flexomagnetism, which was defined analytically as a periodic function with a Gaussian-shaped enhancements and a reduction at the strain boundaries. For the period in $x\in(0, 10)$ $\mu m$, the saturation magnetization is expressed as,
\begin{equation} \label{eq:saturation_set_up}
M_z(x) = M_z^0 + A (e^{-\frac{(x-2.3\mu m)^2}{2\sigma^2}} - e^{-\frac{(x-7.3\mu m)^2}{2\sigma^2}})
\end{equation}
where $A=10^6 \ A/m$ is the amplitude of the Gaussian enhancements and reductions,
$M_z^0 = 2\times10^6  \ A/m$ is a uniform background to ensure that the direction of $\mathbf{M}$ is along the positive $z$-direction, due to the high out-of-plane magnetic field applied during the MFM experiment. $\sigma = 0.5 \ \mu m$ is the standard deviation that suggests a half width of about 1 $\mu m$ for the Gaussian enhancements and reductions. Eq.\ref{eq:saturation_set_up} together with the periodic boundary condition along $x$ result in the alternating modulation in magnetization that is similar to the experimentally measured strain gradient profile shown in Fig.\ref{MFM}(h). The corresponding MFM contrast is evaluated using the built-in MFM module, which computes the out-of-plane derivative of the magnetic stray field generated by the sample's magnetization along the z-axis, convolved with an effective magnetic tip moment modeled as a point monopole at the apex. The MFM lift height is set to be 1 $\mu m$ above the sample. The MFM signal is plotted in a normalized scale.

\section{Acknowledgments}
Magnetic and structural measurements (T.S., Z. L., J.K.K.), density functional theory calculations (Y.T. and K.M.R.), and micromagnetic/structural simulations (J.W., J.-M.H.) were primarily supported by the National Science Foundation through the University of Wisconsin Materials Research Science and Engineering Center award number DMR-2309000. 
GdAuGe film synthesis was supported by the U.S. Department of Energy, Office of Science, Basic Energy Sciences, under award no. DE-SC0023958 (T.S., Z. L., J.K.K.). Additional magnetometry (Z.L., J.K.K.) was supported by the Air Force Office of Scientific Research (FA9550-21-0127). Graphene synthesis by CVD was supported by the U.S. Department of Energy, Office of Science, Basic Energy Sciences, Grant DE-SC0016007 (K.S. and M.S.A.). J.K.K. acknowledges support from the Gordon and Betty Moore Foundation, grant DOI 10.37807/GBMF13808, for revisions of this manuscript. 

The MFM measurements at Rutgers were supported by the Office of Basic Energy Sciences, Division of Materials Sciences and Engineering, U.S. Department of Energy under Award No. DE-SC0018153.

A-H. C., S.K., D.M., M.B., and Z.W. were supported by US Department of Energy, Office of Science, Office of Basic Energy Sciences, Materials Sciences and Engineering Division (helium implantation, some structural measurements and lithography).
Ion irradiation support was conducted as part of a user project at the Center for Nanophase Materials Sciences (CNMS), which is a US Department of Energy, Office of Science User Facility at Oak Ridge National Laboratory.
%Experiments were conducted as part of a user project at the Center for Nanophase Materials Sciences (CNMS), which is a US Department of Energy, Office of Science User Facility at Oak Ridge National Laboratory (some structural measurements and support for helium implantation).

T.B. was supported by the U.S. Department of Energy (DOE) through the University of Minnesota Center for Quantum Materials, under Grant No. DE-SC-0016371. 

This research used the HXN beamline of the National Synchrotron Light Source II, a U.S. Department of Energy (DOE) Office of Science User Facility operated for the DOE Office of Science by Brookhaven National Laboratory under Contract No. DE-SC0012704.
Preliminary nanobeam diffraction measurements were performed at the Center for Nanoscale Materials and Advanced Photon Source, both U.S. Department of Energy Office of Science User Facilities, was supported by the U.S. DOE, Office of Basic Energy Sciences, under Contract No. DE-AC02-06CH11357

\section{Competing interests} The authors do not have competing interests.

\section{Data availability} All data are available in the main text and in the Supplementary materials. 

\bibliographystyle{apsrev}
\bibliography{ref}

\end{document}